\documentclass[aps,twocolumn,reprint,amsmath,amssymb,prb]{revtex4-1}

\usepackage{graphicx,color}
\usepackage{dcolumn}
\usepackage{hyperref}
\usepackage{bm}
\usepackage{stmaryrd}
\usepackage{latexsym}
\usepackage{amssymb}
\usepackage{amsfonts}
\usepackage{amsmath}
\usepackage{epstopdf}

\begin{document}

\title{Spontaneous rotational symmetry breaking in KTaO$_3$ heterointerface superconductors}

\author{Guanqun Zhang,$^{1}$ Lijie Wang,$^{1}$ Jinghui Wang,$^{2}$ Guoan Li,$^{3}$ Guangyi Huang,$^{1}$ Guang Yang,$^{3}$ Huanyi Xue,$^{1}$ Zhongfeng Ning,$^{1}$ Yueshen Wu,$^{2}$ Jin-Peng Xu,$^{3}$ Yanru Song,$^{4,*}$ Zhenghua An,$^{1,5}$ Changlin Zheng,$^{1}$ Jie Shen,$^{3,6*}$ Jun Li,$^{2,*}$ Yan Chen,$^{1}$ and Wei Li$^{1,*}$ }

\affiliation
{$^1$State Key Laboratory of Surface Physics and Department of Physics, Fudan University, Shanghai 200433, China\\
 $^2$School of Physical Science and Technology, ShanghaiTech University, Shanghai 201210, China\\
 $^3$Beijing National Laboratory for Condensed Matter Physics and Institute of Physics, Chinese Academy of Sciences, Beijing 100190, China\\
 $^4$ShanghaiTech Quantum Device Lab, ShanghaiTech University, Shanghai 201210, China\\
 $^5$Institute for Nanoelectronic Devices and Quantum Computing, Fudan University, Shanghai 200433, China\\
 $^6$Songshan Lake Materials Laboratory, Dongguan 523808, China
 }

\date{\today}


\begin{abstract}
Broken symmetries play a fundamental role in superconductivity and influence many of its properties in a profound way. Understanding these symmetry breaking states is essential to elucidate the various exotic quantum behaviors in non-trivial superconductors. Here, we report an experimental observation of spontaneous rotational symmetry breaking of superconductivity at the heterointerface of amorphous (a)-YAlO$_3$/KTaO$_3$(111) with a superconducting transition temperature of 1.86 K. Both the magnetoresistance and superconducting critical field in an in-plane field manifest striking twofold symmetric oscillations deep inside the superconducting state, whereas the anisotropy vanishes in the normal state, demonstrating that it is an intrinsic property of the superconducting phase. We attribute this behavior to the mixed-parity superconducting state, which is an admixture of \emph{s}-wave and \emph{p}-wave pairing components induced by strong spin-orbit coupling inherent to inversion symmetry breaking at the heterointerface of a-YAlO$_3$/KTaO$_3$. Our work suggests an unconventional nature of the underlying pairing interaction in the KTaO$_3$ heterointerface superconductors, and brings a new broad of perspective on understanding non-trivial superconducting properties at the artificial heterointerfaces.
\end{abstract}

\maketitle

The study of heterointerface superconductivity has been a central theme in condensed matter physics communities~\cite{Ref1}. Due to the presence of inversion symmetry breaking and the particular interactions found at their interface between two constitute materials, the strong interplay between the electrons with Coulomb interaction and the interfacial electron-phonon coupling gives rise to novel superconducting behaviors, providing an ideal platform for understanding the underlying rich physical properties and developing the next-generation quantum technologies~\cite{Ref2,Ref3,Ref4}. The archetypal heterointerface superconductivity has been experimentally observed at the heterointerface of crystalline (c)-LaAlO$_3$/SrTiO$_3$ with a superconducting transition temperature ($T_c$) of 250 mK~\cite{Ref5}, which ignites the first fire in heterointerface superconductivity research. Strikingly, a variety of appealing quantum phenomena has also been revealed at the superconducting SrTiO$_3$ heterointerfaces, such as the coexistence of ferromagnetism and superconductivity~\cite{LLi2011,Bert2011,Dikin} and the gate-tunable superconductivity~\cite{ADCaviglia,JBiscaras,Biscaras2012,Zegrodnik,Zegrodnik2021}, indicative of a possible unconventional and non-trivial superconducting phase as the ground state~\cite{Michaeli}. Unfortunately, the extremely low $T_c$ of SrTiO$_3$ heterointerface superconductors is a critical challenge, preventing extensive attentions to further unveil the origin of these emergent quantum phases.

Very recently, unexpected superconductivity is experimentally observed at the heterointerface between polycrystalline (p)-EuO [or amorphous (a)-LaAlO$_3$] and KTaO$_3$ single-crystal substrates which shows a $T_c$ $\sim$ 2 K~\cite{Ref11,ChenScience2021}, approximately one order of magnitude higher than that of c-LaAlO$_3$/SrTiO$_3$~\cite{Ref5}, evoking an exciting opportunity to study the physical properties of heterointerface superconductivity. Although KTaO$_3$ shares many common features with SrTiO$_3$~\cite{Ref11,ChenScience2021,Thompson,Ref14}, the superconductivity of KTaO$_3$ heterointerfaces behaves in a quite different manner. Remarkably, the superconductivity of these heterointerfaces exhibits a strong dependence on the KTaO$_3$ crystalline orientations by compared to the SrTiO$_3$ crystalline orientation independence of superconductivity~\cite{STO110,STO111,Ref13,WangZ2014,PaiYY2018,LiuC2022}. Furthermore, considering the fact that the strong spin-orbit coupling associated with the heavy Ta in 5$d$ orbitals of KTaO$_3$ heterointerfaces is comparable to the bandwidth and the accompanying strong on-site Coulomb repulsion~\cite{KHu2016,Ref37,Venditti}, the combination of strong spin-orbit coupling and the electron-electron interaction is theoretically expected to result in an unconventional superconductivity, including a mix of spin-singlet and spin-triplet components~\cite{LFu2015} as a manifestation of rotational symmetry breaking. Experimentally, an indication of strong in-plane anisotropic electrical resistance in the normal state 
has been reported at the ferromagnetic heterointerface of p-EuO/KTaO$_3$, implying a possible existence of ``strip''-like superconducting phase~\cite{Ref11}. This anisotropy, however, is alternatively attributed to be an extrinsic property of the ferromagnetic p-EuO in theory~\cite{Ref18}, leading to that a consensus on the rotational symmetry breaking of superconductivity in KTaO$_3$ heterointerface superconductors remains elusive. 

\begin{figure*}
\centering
\includegraphics[bb=35 340 540 800,width=9.6cm,height=9cm]{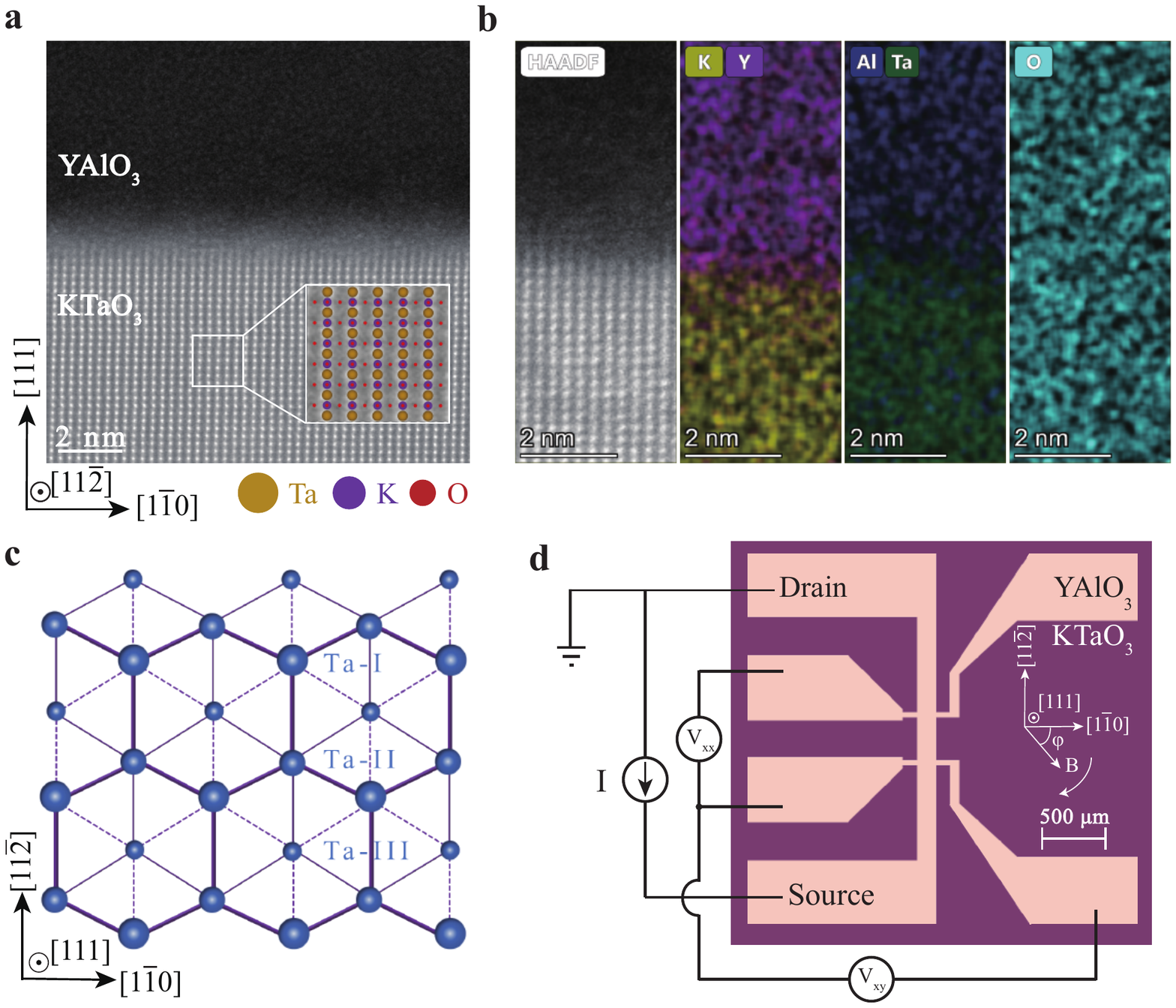}
\caption{\textbf{Structural and composition characterizations of a-YAlO$_3$/KTaO$_3$(111)}. \textbf{a} HAADF-STEM image of a-YAlO$_3$/KTaO$_3$ viewed along the $[11\bar{2}]$ zone axis. The inset shows the enlarged HR-STEM image of KTaO$_3$ overlapped with atomic configuration (colored). \textbf{b} HR-STEM image and the corresponding EDX elemental mapping of interface. \textbf{c} Distribution of Ta$^{5+}$ ions along the $[111]$ crystal axis of KTaO$_3$(111) surface. Ta$^{5+}$ ions are shown with progressively smaller sizes in the three adjacent (111) planes, which are labeled as Ta-I, Ta-II, and Ta-III, respectively. \textbf{d} Hall bar configuration on a-YAlO$_3$/KTaO$_3$(111) heterostructure. Here, $\varphi$ is defined as the in-plane azimuthal angle between the applied magnetic field B and the $[1\bar{1}0]$-axis of the lattice, shown in the inset of (d).
}\label{fig1}
\end{figure*}

Here, we carry out an experimental study on nonmagnetic a-YAlO$_3$ thin films with a wide energy gap of 7.9 eV grown on the polar KTaO$_3$(111) single-crystal substrates. This energy gap is significantly larger than that of a-LaAlO$_3$ (5.6 eV)~\cite{Ref19}, enabling strong confinement potential to restrict the interfacial conducting electrons to a thinner interfacial layer, thus prompting an intriguing quantum behaviors at their interface~\cite{Ref20}. Electrical transport measurements on the as-grown films reveal two-dimensional superconductivity with a $T_c$ of 1.86 K, and a superconducting layer thickness of 4.5 nm. By tuning the in-plane azimuthal angle $\varphi$-dependent magnetic field, both the magnetoresistance and superconducting critical field display pronounced twofold symmetric oscillations deep inside the superconducting state, whereas they vanish in the normal state. These results unambiguously demonstrate that the anisotropy with in-plane rotational symmetry breaking is an intrinsic property of the superconducting phase in a-YAlO$_3$/KTaO$_3$. Through group theory study, we thus classify the inversion symmetry breaking KTaO$_3$ heterointerface superconductors as a mixed-parity unconventional superconductivity with an admixture of $s$-wave and $p$-wave pairing components, a candidate platform for realizing Majorana modes~\cite{PALee2011}.

\begin{figure*}
\centering
\includegraphics[bb=145 65 435 400,width=8.6cm,height=9.5cm]{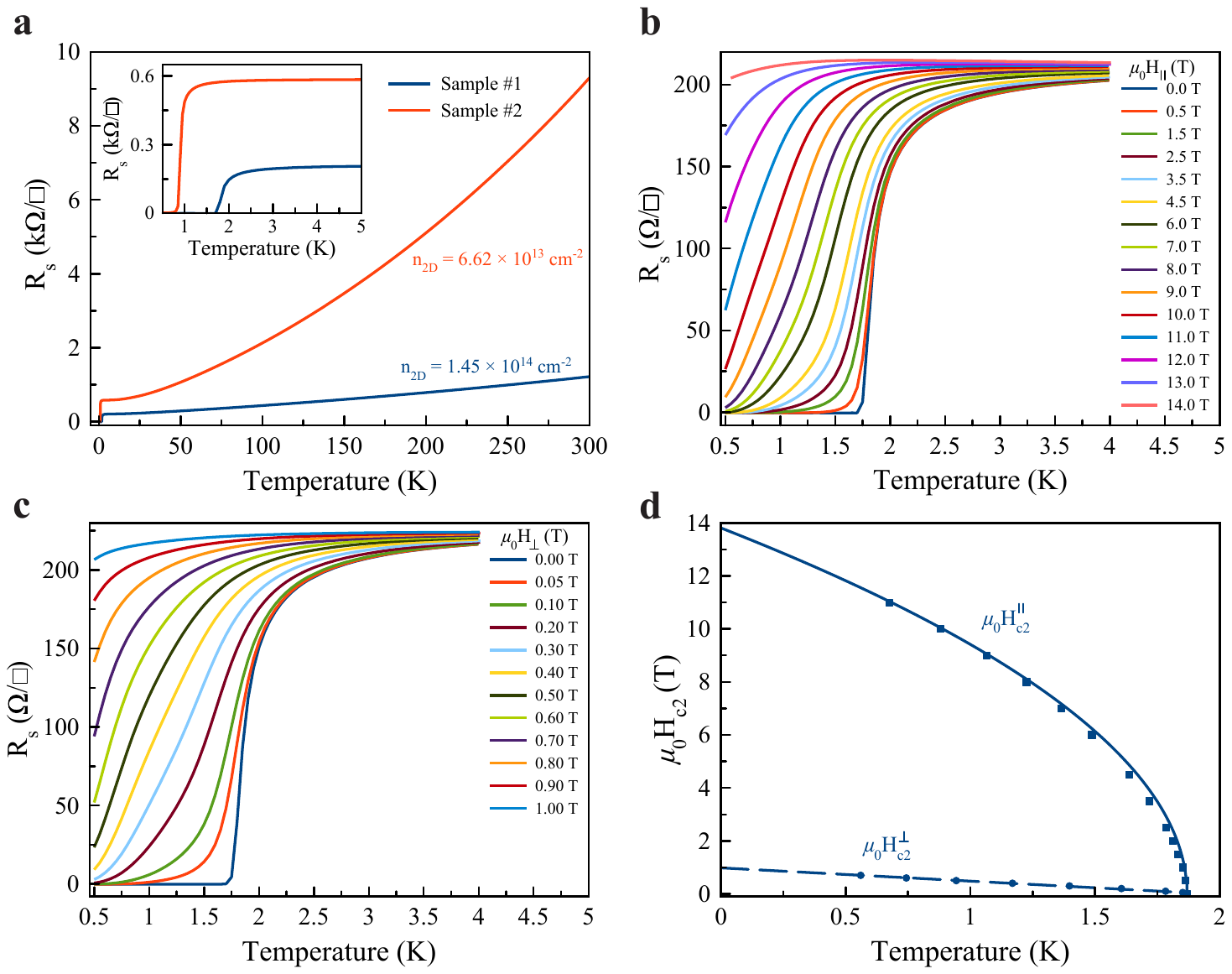}
\caption{\textbf{Superconducting properties of a-YAlO$_3$/KTaO$_3$(111)}. \textbf{a} Electrical resistance (R$_{\mathrm{s}}$) as a function of temperature at zero magnetic field for two representative a-YAlO$_3$/KTaO$_3$ heterostructures (Samples \#1 and \#2). Low temperature dependence of R$_{\mathrm{s}}$ is illustrated in the inset of (a). Magnetoresistance for fields \textbf{b} parallel and \textbf{c} perpendicular to the plane surface of Sample \#1. \textbf{d} Temperature dependence of the upper critical field $\mu_0$H$_{c2}$ ($\mu_0$H$_{c2}^{\parallel}$ for the in-plane field along the $[11\bar{2}]$-axis shown in Fig.~\ref{fig1}d and $\mu_0$H$_{c2}^{\perp}$ for the out-of-plane field).}
\label{fig2}
\end{figure*}

\notag\

\noindent\textbf{Results}

\noindent The a-YAlO$_3$/KTaO$_3$ heterostructures are prepared by depositing a-YAlO$_3$ films on (111)-oriented KTaO$_3$ single-crystal substrates using pulsed laser deposition. Atomic force microscopy characterizations show that the surface of KTaO$_3$ substrates and a-YAlO$_3$ films are atomically flat (see Supplementary Fig. 1). X-ray diffraction confirms the absence of epitaxial peaks of YAlO$_3$ (see Supplementary Fig. 2 and Supplementary Fig. 3), thus suggesting that the YAlO$_3$ film is not in a well-defined crystalline phase. The microstructure of the interface is further examined by aberration-corrected scanning transmission electron microscopy (STEM). From the high angle annular dark field (HAADF)-STEM image shown in Fig.~\ref{fig1}a, it can be seen that the homogeneous and amorphous phase YAlO$_3$ thin film is grown on the KTaO$_3$(111) substrate (also see Supplementary Fig. 4). Looking at the sample from a larger field of view, the thickness of the a-YAlO$_3$ film is found to be about 60 nm. High-resolution (HR)-STEM imaging shown in Fig.~\ref{fig1}a and Supplementary Fig. 4, and energy dispersive X-ray spectroscopy (EDX) elemental mapping shown in Fig.~\ref{fig1}b indicate that the abrupt and smooth interface between KTaO$_3$ single-crystal substrate and a-YAlO$_3$ film is resolved structurally and chemically. These results are consistent with previous studies on a-LaAlO$_3$/KTaO$_3$(111)~\cite{Ref11,ChenScience2021}, a-LaAlO$_3$/KTaO$_3$(110)~\cite{Ref13}, a-LaAlO$_3$/KTaO$_3$(001)~\cite{Zhang001}, and a-AlO$_x$/KTaO$_3$(111)~\cite{Mallik2022}.

Figure~\ref{fig2}a shows the temperature-dependent sheet resistance R$_{\mathrm{s}}$ on two representative as-grown a-YAlO$_3$ thin films (Samples \#1 and \#2 with growth temperatures of 780 and 650 $^{\circ}$C, respectively) with the Hall bar configuration, schematically illustrated in Fig.~\ref{fig1}d. A typical metallic behavior is visible in a wide temperature range, indicating that a two-dimensional electron gas is formed at their interface induced by a candidate mechanism of the formation of oxygen vacancies at the surface of KTaO$_3$~\cite{Ref13,Ning2023}, similar to that in the sister a-LaAlO$_3$/SrTiO$_3$~\cite{LiuZQ}. The transverse Hall resistance R$_{\mathrm{xy}}$ is obtained from Hall measurements at 5 K, and reveals that the charge carriers in the a-YAlO$_3$/KTaO$_3$ are electrons. The estimated carrier density is about 1.45$\times$10$^{14}$ and 6.62$\times$10$^{13}$ cm$^{-2}$ for Samples \#1 and \#2, respectively. The electron mobility for Samples \#1 and \#2 is thus 193.6 and 159.7 cm$^2$V$^{-1}$s$^{-1}$. These results are highly universal and reproducible (see Supplementary Fig. 5, Supplementary Note 1, and Supplementary Table 1) and reasonably consistent with previous electrical transport studies on the KTaO$_3$ heterointerfaces~\cite{Ref11,Ref13}. Remarkably, as the temperature is further decreased, the resistance R$_{\mathrm{s}}$ undergoes a narrow and sharp transition with a transition width of less than 0.5 K to a zero-resistance state, signaling the appearance of superconductivity at the heterointerface of a-YAlO$_3$/KTaO$_3$. The critical temperature is determined to be $T_c$ = 1.86 and 0.92 K for Samples \#1 and \#2, respectively, as defined by where the resistance is at the midpoint of the normal electrical resistance at 5 K, i.e. R$_{\mathrm{s}}$($T_c$) = 0.5$\times$R$_{\mathrm{s}}$(5 K).

\begin{figure*}
\centering
\includegraphics[bb=155 155 430 400,width=8.6cm,height=7.8cm]{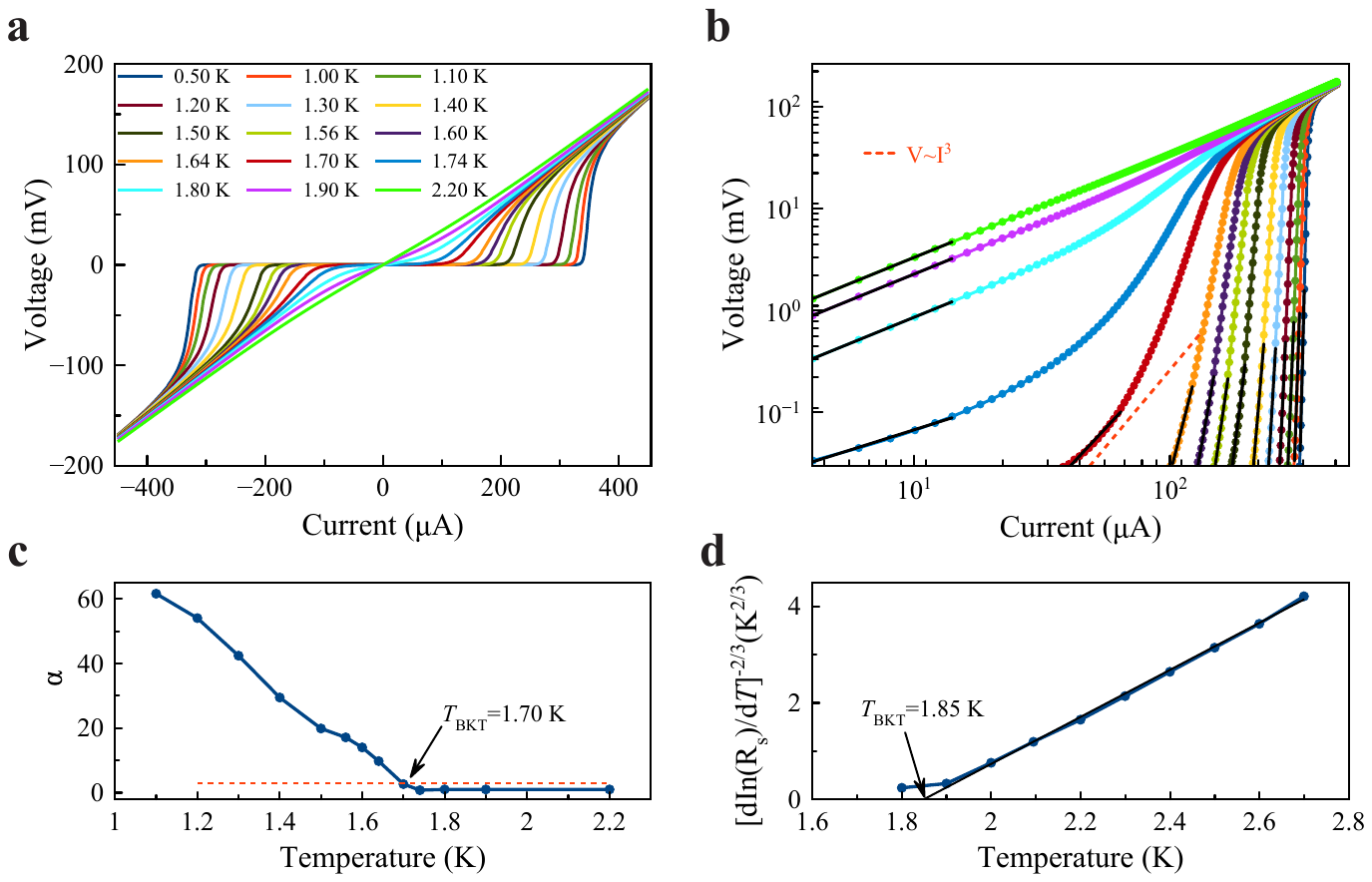}
\caption{\textbf{Two-dimensional superconducting behavior of a-YAlO$_3$/KTaO$_3$(111)}. \textbf{a} Temperature-dependent $\mathrm{I}$-$\mathrm{V}$ measurements (Sample \#1). \textbf{b} Corresponding logarithmic scale representation of (a). The long red dashed line denotes the $\mathrm{V}\sim \mathrm{I}^3$ dependence. \textbf{c} Temperature dependence of the power-law exponent $\alpha$, as deduced from the fits shown in (b). \textbf{d} R$_{\mathrm{s}}(T)$ dependence of the same sample, plotted on a $[\mathrm{d}\ln(\mathrm{R}_{\mathrm{s}})/\mathrm{d}T]^{-2/3}$ scale.}
\label{fig3}
\end{figure*}

To further characterize the superconducting behaviors in a-YAlO$_3$/KTaO$_3$, we measure the magnetoresistance R$_{\mathrm{s}}$($\mu_0$H) (here, $\mu_0$ is the vacuum permeability) at various temperatures between 0.5 and 5 K with fields parallel ($\mu_0$H$_{\parallel}$) and perpendicular ($\mu_0$H$_{\perp}$) to the plane surface of Sample \#1, as shown in Fig.~\ref{fig2}b,c, respectively. The fundamental superconducting behavior is clearly perceived. Indeed, the magnetoresistance R$_{\mathrm{s}}$($\mu_0$H) varies differently with $\mu_0$H$_{\parallel}$ and $\mu_0$H$_{\perp}$, and both the upper critical fields $\mu_0$H$_{c2}^{\parallel}$ and $\mu_0$H$_{c2}^{\perp}$ parallelly shift to a lower value with the increase of the temperature, where $\mu_0$H$_{c2}$ are evaluated at the midpoints of the normal state resistance at 5 K. These results provide an indication of a two-dimensional superconducting feature in a-YAlO$_3$/KTaO$_3$. The temperature-dependent upper critical fields $\mu_0$H$_{c2}$ are summarized in Fig.~\ref{fig2}d and are well fitted by the phenomenological two-dimensional Ginzburg-Landau (G-L) model~\cite{Ref21}: $\mu_0$H$_{c2}^{\perp}(T)=\frac{\Phi_0}{2\pi\xi^2_{\mathrm{GL}}}(1-\frac{T}{T_c})$ and $\mu_0$H$_{c2}^{\parallel}(T)=\frac{\Phi_0\sqrt{12}}{2\pi\xi_{\mathrm{GL}}d_{\mathrm{SC}}}\sqrt{1-\frac{T}{T_c}}$, where $\Phi_0$, $\xi_{\mathrm{GL}}$, and $d_{\mathrm{SC}}$ denote a flux quantum, the in-plane superconducting coherence length at $T$ = 0 K, and the effective thickness of superconductivity, respectively. Using the extrapolated $\mu_0$H$_{c2}^{\perp}(0)$ = 0.98 T and $\mu_0$H$_{c2}^{\parallel}(0)$ = 13.81 T, we find $\xi_{\mathrm{GL}}$ = 18.4 nm and $d_{\mathrm{SC}}$ = 4.5 nm, where $\xi_{\mathrm{GL}}$ is significantly larger than $d_{\mathrm{SC}}$, suggesting a two-dimensional nature of superconductivity. Additionally, the in-plane $\mu_0$H$_{c2}^{\parallel}(0)$ is substantially larger than the Pauli-paramagnetic pair-breaking field $\mathrm{B}_\mathrm{P}\approx$ 3.46 T based on the BCS theory in the weak-coupling limit~\cite{Ref22,Ref23}. High values of $\mu_0$H$_{c2}^{\parallel}(0)$ excessing $\mathrm{B}_\mathrm{P}$ could be realized in the presence of strong spin-orbit coupling owing to the elastic scattering, which results in the suppression of spin paramagnetism effects. The violation of this paramagnetic limit is a common phenomenon in heterointerface superconductors~\cite{Ref11,Ref24}, especially when the superconducting layer thickness is in the range $d_{\mathrm{SC}}<$ 20 nm. However, the underlying mechanism for realizing $\mu_0$H$_{c2}^{\parallel}(0)$ value in excess of $\mathrm{B}_\mathrm{P}$ remains an open question~\cite{Ref11}. Furthermore, the thickness of the superconducting layer in a-YAlO$_3$/KTaO$_3$(111) is approximately estimated as thin as $d_{\mathrm{SC}}$ = 4.5 nm based on the framework of the phenomenological two-dimensional G-L model~\cite{Ref21}. This result could be intuitively expected, since the strong confinement potential induced by YAlO$_3$ significantly restricts the superconducting electrons to a thinner superconducting layer~\cite{Ref20}. On the other hand, the out-of-plane polar angle $\theta$-dependent upper critical field H$_{c2}^{\theta}$ at 1.5 K quantitatively verifies the behavior expected from a two-dimensional structure in a-YAlO$_3$/KTaO$_3$, as shown in Supplementary Fig. 6. The $\theta$-dependent $\mu_0$H$_{c2}^{\theta}$ are quantitatively fitted by the two-dimensional Tinkham formula and the three-dimensional anisotropic G-L model, given by $\frac{\mathrm{H}_{c2}^{\theta}|\cos\theta|}{\mathrm{H}_{c2}^{\perp}} + (\frac{\mathrm{H}_{c2}^{\theta}\sin\theta}{\mathrm{H}_{c2}^{\parallel}})^2 = 1$ and $(\frac{\mathrm{H}_{c2}^{\theta}\cos\theta}{\mathrm{H}_{c2}^{\perp}})^2 + (\frac{\mathrm{H}_{c2}^{\theta}\sin\theta}{\mathrm{H}_{c2}^{\parallel}})^2 = 1$, respectively~\cite{Ref29,Ref31}. A cusp-like peak is clearly observed at $\theta=90^{\circ}$ (see Supplementary Fig. 6), which is well described by the two-dimensional Tinkham model, as frequently observed in heterointerface superconductivity~\cite{Ref31,ZhangGQ2022} and layered transition metal dichalcogenides~\cite{Ref30,Ref32}.

\begin{figure*}
\centering
\includegraphics[bb=150 100 430 410,width=8.8cm,height=9cm]{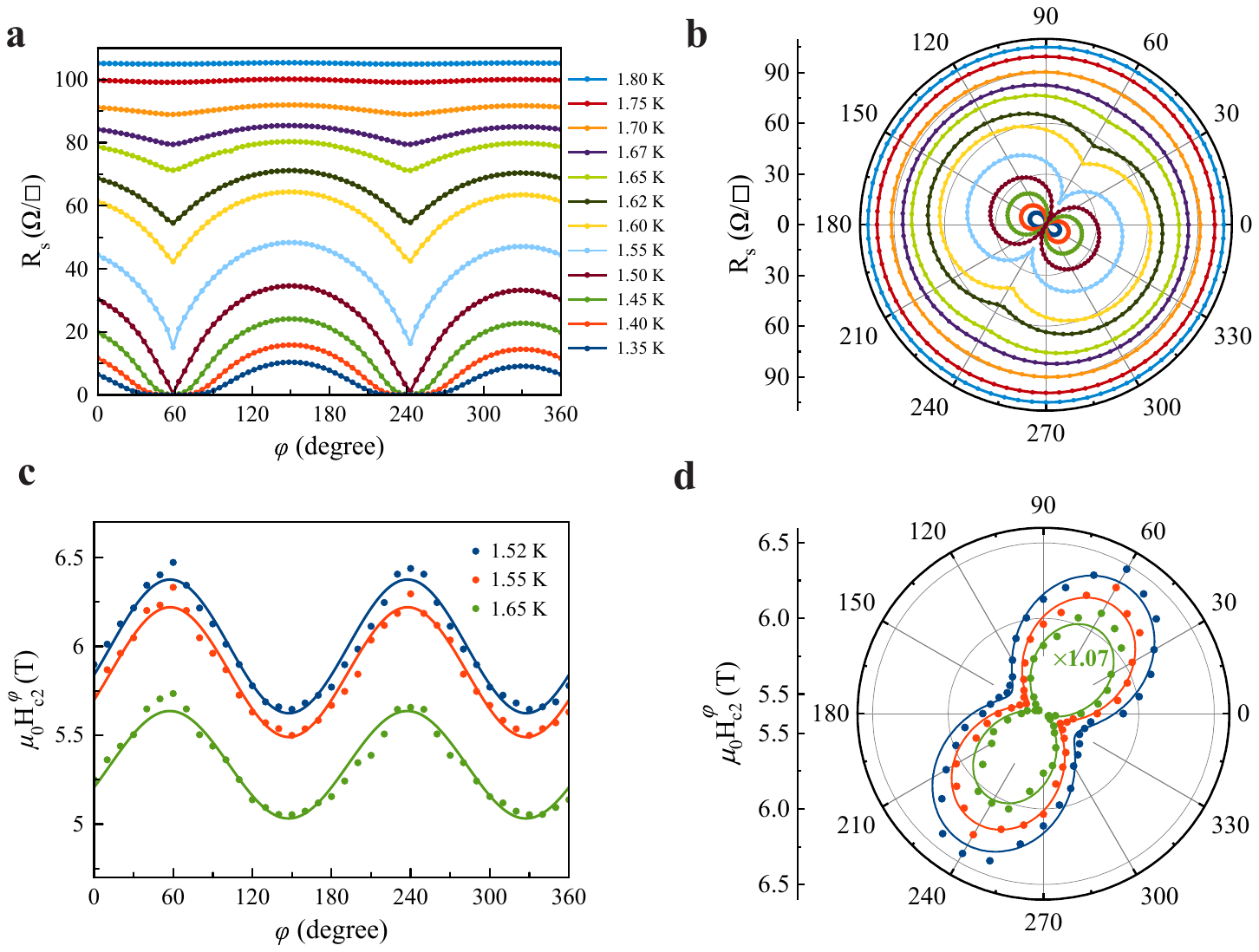}
\caption{\textbf{In-plane twofold symmetric oscillations in a-YAlO$_3$/KTaO$_3$(111)}. \textbf{a} In-plane angular-dependent magnetoresistance R$_{\mathrm{s}}$ at various temperatures for an applied field of 1 T. \textbf{b} Polar plots of the data in (a). \textbf{c} In-plane angular-dependent $\mu_0$H$^{\varphi}_{c2}$ at various temperatures. \textbf{d} Polar plots of the data in (c). The solid lines in (c) and (d) are the theoretical fits of the H$^{\varphi}_{c2}$ using the gap function of $|\Delta_{\mathrm{gap}}|^2$ with an admixture of $s$-wave and $p$-wave pairings, $\Delta_{\mathrm{gap}}(\mathbf{k})=i[\Delta_s\hat{\sigma}_0 + \Delta_p\mathrm{sin}(k_x)\hat{\sigma}_3]\hat{\sigma}_2$. Here, $\Delta_s$ and $\Delta_p$ are the pairing amplitudes of $s$-wave and $p$-wave, respectively, $\hat{\sigma}$ is the vector of Pauli matrices, and $\mathbf{k}$ is the momentum.}
\label{fig4}
\end{figure*}

Since the superconductivity in a-YAlO$_3$/KTaO$_3$ is two-dimensional, the Berezinskii-Kosterlitz-Thouless (BKT) transition describes superconducting phase coherence~\cite{Ref33,Ref34}. Here, the BKT transition temperature defines the vortex unbinding transition, and can be determined using current-voltage ($\mathrm{I}$-$\mathrm{V}$) measurements as a function of temperature $T$, as shown in Fig.~\ref{fig3}a. Below $T_c$, we find a critical current $\mathrm{I}_c$, whose value decreases with increase in temperature. The maximal value of $\mathrm{I}_c$ is $\sim$330 $\mu$A at 0.5 K, which is substantially larger than that previously observed in the KTaO$_3$ heterointerfaces~\cite{Ref11,Ref13}. Such a high critical current value probably originates from the high charge carrier concentration (about 1.45$\times$10$^{14}$ cm$^{-2}$, Sample \#1 in Fig.~\ref{fig2}a) confined in a thinner superconducting layer of a-YAlO$_3$/KTaO$_3$, promising for large-scale applications in superconductor-based devices. In Fig.~\ref{fig3}b, we also plot the characteristics $\mathrm{I}$-$\mathrm{V}$ on a log-log scale, and observe that the slope of the $\mathrm{I}$-$\mathrm{V}$ curve smoothly evolves from the normal ohmic state, $\mathrm{V}\propto \mathrm{I}$, to a steeper power law resulting from the current exciting free-moving vortices, $\mathrm{V}\propto \mathrm{I}^{\alpha(T)}$, with $\alpha(T_{\mathrm{BKT}})$ = 3. In Fig.~\ref{fig3}c, a value $T_{\mathrm{BKT}}$ = 1.7 K is interpolated, which is consistent with $T_c$ as defined in Fig.~\ref{fig2}a. In addition, close to $T_{\mathrm{BKT}}$, an R$_{\mathrm{s}} = R_0 \exp[-b(T/T_{\mathrm{BKT}}-1)^{-1/2}]$ dependence, where $R_0$ and $b$ are material parameters, is expected~\cite{Ref35}. As shown in Fig.~\ref{fig3}d, the measured R$_{\mathrm{s}}(T)$ is also consistent with this behavior and yields $T_{\mathrm{BKT}}$ = 1.85 K, in good agreement with the analysis of the $\alpha$ exponent shown in Fig.~\ref{fig3}c.

Next, we turn to discuss the in-plane anisotropy of superconductivity in a-YAlO$_3$/KTaO$_3$ using an in-plane azimuthal angle $\varphi$-dependent magnetoresistance, where $\varphi$ is defined as the azimuthal angle between the magnetic field and the $[1\bar{1}0]$-axis of the lattice, as indicated in Fig.~\ref{fig1}d. Care has been taken to rule out the inevitable misalignment effects of an accidental out-of-plane component of the field, when the vector magnet is utilized. In the normal state ($T$ = 1.8 K in Fig.~\ref{fig4}a) of Sample \#4 (Supplementary Fig. 7), the magnetoresistance R$_{\mathrm{s}}$ is found to be essentially independent of $\varphi$, displaying an isotropic behavior. Whereas in the superconducting state ($T$ = 1.5 K in Fig.~\ref{fig4}a), we observe a pronounced twofold symmetric oscillations of the R$_{\mathrm{s}}$ (see Fig.~\ref{fig4}b), which is consistent across multiple samples including the heterointerfaces of a-YAlO$_3$/KTaO$_3$(111) and sister a-LaAlO$_3$/KTaO$_3$(111) (see Supplementary Fig. 8 $-$ Supplementary Fig. 10 and Supplementary Note 2). In this case, the anisotropic R$_{\mathrm{s}}$ attains the maximum value when the magnetic field is directed along the special $[1\bar{2}1]$-axis ($\varphi =-30^{\circ}$ or 150$^{\circ}$) that is in the direction of one of the principal axes of KTaO$_3$(111) shown in Fig.~\ref{fig1}c, and becomes minimum when the position with respect to that of maximum is shifted by 90$^{\circ}$ ($\varphi = 60^{\circ}$ or 240$^{\circ}$). This finding implies that an extrinsic error from the experimental setup is unlikely to the source of the observed twofold anisotropy of magnetoresistance in the superconducting state at the heterointerface of a-YAlO$_3$/KTaO$_3$. In particular, the significantly large anisotropic ratio of R$_{\mathrm{s}}$($\varphi$=60$^{\circ}$)/R$_{\mathrm{s}}$($\varphi$=150$^{\circ}$) = 0.03 at 1.5 K corresponds to a putative misalignment angle estimated up to 88.03$^{\circ}$ ($\cos$ 88.03$^{\circ}$ = 0.03) between the field and the basal plane~\cite{XiangY2021}, which is impossible for such a large angle misalignment in the vector magnet, we could exclude the possible contribution from an accidental misalignment of the field with the film plane. Since the magnetoresistance minima approach zero in R$_{\mathrm{s}}$($\varphi$) curve measured at 1.5 K (see Fig.~\ref{fig4}a), and considering the fact that the existence of conspicuous twofold symmetry in magnetoresistance manifests deep inside the superconducting region, which vanishes in the normal state (see Fig.~\ref{fig4}a and Supplementary Fig. 11), we could further rule out the possibilities of extrinsic contributions, such as the magnetic field induced Lorentz force effect~\cite{Ref36} and the Fermi surface inherent to the KTaO$_3$ with respect to the underlying threefold lattice symmetry~\cite{Ref37} (Fig.~\ref{fig1}c and Supplementary Fig. 3), and thus demonstrate that this anisotropy with rotational symmetry breaking is an intrinsic property of the superconducting phase in a-YAlO$_3$/KTaO$_3$.

To further reveal the twofold symmetric superconductivity in a-YAlO$_3$/KTaO$_3$ in terms of the superconducting gap structure, we extract the upper critical field $\mu_0$H$_{c2}$ from the $\varphi$-dependent magnetoresistance R$_{\mathrm{s}}$ in the superconducting region determined by the criterion of 90\% sheet resistance dropped from normal state, as shown in Fig.~\ref{fig4}c. Here, it should be pointed out that although the values of H$_{c2}$ are changed by different criteria, the symmetry of H$_{c2}$ itself remains qualitatively (see Supplementary Fig. 12). In addition, it should be noted that the data shown here have been taken by averaging the raw data with positive and negative magnetic fields to avoid the possible asymmetric problem. Remarkably, the in-plane $\varphi$-dependent $\mu_0$H$_{c2}^{\varphi}$ also displays twofold symmetric oscillations (see Fig.~\ref{fig4}d), providing additional strong evidence for the twofold rotational symmetry of the superconductivity in a-YAlO$_3$/KTaO$_3$. Furthermore, the oscillation of $\mu_0$H$_{c2}^{\varphi}$ has a $\pi$ phase shift compared with that of the R$_{\mathrm{s}}$ (see Fig.~\ref{fig4}b) such that for the $\varphi$ value where superconductivity is hardest to suppress, $\mu_0$H$_{c2}^{\varphi}$ is the largest and R$_{\mathrm{s}}$ is the lowest (Fig.~\ref{fig4}b,d), as expected from our intuitions~\cite{Ref36,Ref38,LiJun2017}. Since $\mu_0$H$_{c2}^{\varphi}$ achieves its maximum for the field applied perpendicular to the main crystallographic axis, and minimum for the direction along the main crystallographic axis (see Fig.~\ref{fig1}d and Fig.~\ref{fig4}d), the superconducting gap leads to a maximum (or minimum) direction perpendicular (or parallel) to the main crystallographic axis, 
manifesting a rotational symmetry breaking state of superconducting a-YAlO$_3$/KTaO$_3$ with the direction of the minimum gap spontaneously pinned to the main crystallographic axis.

\notag\

\noindent\textbf{Discussion}

\noindent Having experimentally established the intrinsic twofold anisotropy of the superconducting state of a-YAlO$_3$/KTaO$_3$, we now proceed to elaborate about its origin using the underlying symmetries of the crystal structure without requiring the details of the pairing mechanisms based on the group theoretical formulation of the Ginzburg-Landau theory~\cite{Ref40}(also see Supplementary Note 2 and Supplementary Note 3 in details). This allows us to deduce fundamental information about the superconducting ground state in the a-YAlO$_3$/KTaO$_3$ heterointerface superconductors. From the viewpoint of group symmetry, if a superconductor possesses an inversion symmetry, the Pauli principle requires a totally antisymmetric Cooper pair wavefunction, which imposes the condition that the superconducting states should be either spin-singlet or spin-triplet, whereas mixed-parity states are forbidden~\cite{Ref40}. In the a-YAlO$_3$/KTaO$_3$ the lack of inversion symmetry, however, tends to mix spin-singlet and spin-triplet driven by strong spin-orbit coupling~\cite{Ref41}. Indeed, the conducting electrons with strong spin-orbit coupling originating from the heavy Ta $5d$ orbitals has been elucidated at the KTaO$_3$ heterointerfaces~\cite{Zhang001,Ref37,SOC2021,Ref43,Wadehra2020,Trier2022,Gupta_1,Gupta_2} (also see Supplementary Fig. 13). Due to the absence of a mirror plane parallel to the interface of a-YAlO$_3$/KTaO$_3$, the point group of a-YAlO$_3$/KTaO$_3$ is $C_{3v}$, which does not contain the symmetry element of an inversion. This situation is analogue to non-centrosymmetric superconductors~\cite{Ref41,YipS}. Upon inspecting the character table of $C_{3v}$ point group tabulated in Supplementary Table 2, we notice that the mixed-parity superconducting state only belongs to the $A_1$+$E$-representation with the possible basis function of $s$+$p$. Notably, the two-dimensional irreducible representation of $E$ could spontaneously break the threefold rotational symmetry of the crystal (see Fig.~\ref{fig1}c and Supplementary Fig. 3), leading to a subsidiary uniaxial anisotropy or nematic superconductivity~\cite{YPan2016,Ref44}, such as a uniaxial $p_x$-wave or $p_y$-wave pairing. Since the upper critical field is proportional to the square of the superconducting gap amplitude based on the Ginzburg-Landau theory and the Pippard definition of the coherence length~\cite{XiangY2021}, $\mu_0$H$_{c2}^{\varphi}\propto |\Delta_{\mathrm{gap}}(\varphi)|^2$, only the $s$+$p_x$-wave pairing with the gap function of $\Delta_{\mathrm{gap}}(\mathbf{k})=i[\Delta_s\hat{\sigma}_0 + \Delta_p\mathrm{sin}(k_x)\hat{\sigma}_3]\hat{\sigma}_2$ (here, $\Delta_s$ and $\Delta_p$ are the pairing amplitudes of $s$-wave and $p$-wave, respectively, $\hat{\sigma}$ is the vector of Pauli matrices, and $\mathbf{k}$ is the momentum)~\cite{Annett,YipS}, could give rise to an overall twofold anisotropic gap and well reproduce the exact topology of the anisotropic H$^{\varphi}_{c2}$ shown in Fig.~\ref{fig4}d. Therefore, the mix of $s$-wave and $p$-wave pairings driven by strong spin-orbit coupling is suggested to the source of the experimentally observed twofold anisotropic superconductivity at the KTaO$_3$ heterointerfaces, which has long been a topic of interest sought in condensed matter physics. Further experiments, including probes of the superconducting gap by tunneling spectroscopy and/or Josephson junction experiments, will also be helpful for clarifying the underlying mixed-parity pairing nature of the twofold symmetric superconductivity that we observe.

In summary, we have experimentally observed spontaneous rotational symmetry breaking from threefold to twofold in the superconducting state of KTaO$_3$(111) heterointerfaces with respect to an application of in-plane magnetic field. This in-plane anisotropic superconductivity is theoretically attributed to the intrinsic nature of mixed-parity unconventional superconductivity with an admixture of $s$-wave and $p$-wave pairing components, bringing with it fresh new insights into the study of emergent fascinating and non-trivial superconducting properties at the heterointerfaces with inversion symmetry breaking.

\notag\

\noindent\textbf{Methods}

\noindent{\bf Thin film growth and structural characterizations.} a-YAlO$_3$ thin films are grown on KTaO$_3$(111) single-crystal substrates (5$\times$5$\times$0.5 mm$^3$) by pulsed laser deposition in an ultrahigh vacuum chamber (base pressure of 10$^{-9}$ Torr). Prior to growth, the KTaO$_3$ substrates are annealed at 600 $^{\circ}$C for 30 mins in ultrahigh-vacuum to obtain a smooth surface (Supplementary Fig. 1). During deposition, a single crystal YAlO$_3$ target (Kurt J. Lesker Company) is used to grow the a-YAlO$_3$ films with a KrF excimer laser (Coherent 102, wavelength: $\lambda$ = 248 nm). A pulse energy density of 1.5 J/cm$^2$ and a repetition rate of 2 Hz are used. The a-YAlO$_3$ films are deposited at temperatures ranging from 600 to 900 $^{\circ}$C in a vacuum chamber to promote growth of the superconducting phase. All the samples are cooled to room temperature at a constant rate of 20 $^{\circ}$C/min in vacuum after deposition. The quality of a-YAlO$_3$ films under ambient conditions is examined by atomic force microscopy (AFM, Asylum Research MFP-3D Classic) and by four-circle  X-ray diffraction (XRD, Bruker D8 Discover, Cu $\mathrm{K}\alpha$ radiation, $\lambda$ = 1.5406 \AA) operated in HR mode using a three-bounce symmetric Ge (022) crystal monochromator.

\notag\

\noindent{\bf STEM measurements.} Cross-sectional specimens for electron microscopy are prepared with Focused Ion Beam (FIB) (Helios-G4-CX, Thermo Fisher Scientific) using lift-out method. The HR-STEM images are performed on a double aberration corrected field-emission STEM (Themis Z, Thermo Fisher Scientific) operated at 300 kV. For HAADF-STEM imaging, the semi-convergent angle of the probe forming lens is set to 21.4 mrad. The geometric aberrations within the probe forming lens aperture have been effectively tuned to zero using probe corrector (SCORR, CEOS GmbH). The semi-collection angle of the HAADF detector is $76-200$ mrad. Furthermore, the chemical composition of the interface is qualitatively analyzed using EDX in STEM spectrum imaging mode. The EDX are collected using 4 silicon drift detector (SDD) system (Super X detector, Thermo Fisher Scientific). The beam current for STEM-EDX analysis is about 200 pA.

\notag\

\noindent{\bf Electrical transport measurements.} The electrical transport measurements are performed using a commercial cryostat with temperature ranging from 1.5 to 300 K (Oxford Instruments TeslatronPT cryostat system), physical properties measurement system with temperature ranging from 0.5 to 300 K (PPMS, Quantum Design), and 10 mK dilution refrigerator with vector magnet (Oxford Instruments Triton 200). The Hall bar structure (Fig.~\ref{fig1}d) is fabricated by ion-beam etching to systemically measure the electrical transport properties. The vector magnet is utilized to reveal the in-plane anisotropy of magnetoresistance in the superconducting state shown in Fig.~\ref{fig4}a, and the samples are mounted on a mechanical rotator in a $^4$He cryostat to clarify the anisotropy of H$_{c2}$ shown in Fig.~\ref{fig4}c. The misalignment of the field with the film plane is estimated to be less than $2^{\circ}$ and $7^{\circ}$ for vector magnet and mechanical rotator, respectively, as our experimental errors.

\notag\

\noindent\textbf{Data availability}

\noindent The relevant data supporting our key findings are available within the article and the Supplementary Information file. All raw data generated during our current study are available from the corresponding authors upon reasonable request.

\notag\

\noindent\textbf{Acknowledgements}

\noindent
This work is supported by the National Natural Science Foundation of China (Grant Nos. 61871134 and 11927807) and Shanghai Science and Technology Committee (Grant Nos. 23ZR1404600 and 20DZ1100604). The authors thank the Synergetic Extreme Condition User Facility (SECUF) at Institute of Physics, Chinese Academy of Sciences.

\notag\

\noindent\textbf{Author contributions}

\noindent
G.Z., L.W., J.W., and G.L. contributed equally to this work. W.L. conceived the project and designed the experiments. G.Z. grew the samples. G.Z., J.W., G.L., G.Y., Y.S., L.W., Z.N., J.S., and J.L. performed the electrical transport measurements. H.X. and Z.A. fabricated the Hall bar structure on the thin films. G.H. and C.Z. performed scanning transmission electron microscopy measurements. W.L. wrote the paper. All authors discussed the results and gave approval to the final version of the manuscript.

\notag\

\noindent\textbf{Competing interests}

\noindent
The authors declare no competing interests.

\notag\

\noindent\textbf{Additional information}

\noindent
Supplementary information

\notag\

\noindent\textbf{Correspondence and requests} for materials should be addressed to Y. Song, J. Shen, J. Li or W. Li.

\notag\

\noindent $^{*}$To whom correspondence should be addressed. E-mail: songyr@shanghaitech.edu.cn, shenjie@iphy.ac.cn, lijun3@shanghaitech.edu.cn, or w$\_$li@fudan.edu.cn

\end{document}